\documentclass[10pt,twocolumn,amsmath,amssymb,aps,pre,superscriptaddress]{revtex4-2}

\usepackage{graphicx}
\usepackage{lipsum}
\usepackage{xcolor}
\usepackage{ragged2e}
\usepackage{lmodern}
\usepackage{tabularx}
\usepackage[T1]{fontenc}
\newcommand{\dd}{\operatorname{d}}
\usepackage{bm}
\usepackage[separate-uncertainty=true]{siunitx}

\renewcommand{\dd}{\operatorname{d}} 
\newcommand{\ie}{\textit{i.e.}}    
\newcommand{\iu}{\operatorname{i}} 

\begin{document}

\title{Run and tumble dynamics of a soft robotic cell}

\author{S. Mohapatra}
\thanks{These authors contributed equally to this work}
\affiliation{PULS, Institute for Theoretical Physics, FAU Erlangen-Nürnberg, 91058 Erlangen, Germany}

\author{F. W\'ery}
\thanks{These authors contributed equally to this work}
\affiliation{GRASP, Institute of Physics B5a, University of Li\`ege, B4000 Li\`ege, Belgium}

\author{F. Novkoski}
\affiliation{PULS, Institute for Theoretical Physics, FAU Erlangen-Nürnberg, 91058 Erlangen, Germany}
\affiliation{GRASP, Institute of Physics B5a, University of Li\`ege, B4000 Li\`ege, Belgium}

\author{P. Nowakowski}
\affiliation{Group for Computational Life Sciences, Division of Physical Chemistry, Ru\dj{}er Bo\v{s}kovi\'{c} Institute, Zagreb 10000, Croatia}

\author{A.-S. Smith}
\thanks{Correspondence to: smith@physik.fau.de, asmith@irb.hr or nvandewalle@uliege.be}
\affiliation{PULS, Institute for Theoretical Physics, FAU Erlangen-Nürnberg, 91058 Erlangen, Germany}
\affiliation{Group for Computational Life Sciences, Division of Physical Chemistry, Ru\dj{}er Bo\v{s}kovi\'{c} Institute, Zagreb 10000, Croatia}

\author{N. Vandewalle}
\thanks{Correspondence to: smith@physik.fau.de, asmith@irb.hr or nvandewalle@uliege.be}
\affiliation{GRASP, Institute of Physics B5a, University of Li\`ege, B4000 Li\`ege, Belgium}

\begin{abstract}
The continuous regulation of transport properties through softness remains a longstanding challenge in active matter. Here, we show that encasing a programmable active particle within a deformable membrane naturally gives rise to intermittent stop-and-go dynamics, with ballistic motion at short times crossing over to diffusion at long times. Crucially, membrane softness acts as a single control parameter that continuously tunes persistence, 
intermittency, and long-time transport, linking the internal driving to the emergent locomotion of the synthetic cell. Combining experiments, simulations, and a run-and-tumble theoretical framework, we identify the minimal physical ingredients underlying this behavior and establish design principles for programmable soft active transport, opening new avenues at the interface of active matter physics and synthetic robotics.
\end{abstract}

\maketitle

Living systems have long inspired efforts in synthetic active matter to develop physical frameworks for emergent non-equilibrium behavior and to elucidate fundamental biological principles~\cite{marchetti2013hydrodynamics, cates2015motility, bechinger2016active}. For example, motivated by the softness and activity of cells and bacteria, active vesicles enclosing cytoskeleton-inspired assemblies were instrumental for demonstrating the importance of membrane deformability for their internal organization~\cite{gandikota2023rectification, paoluzzi2016shape, chen2017rotational, uplap2023design, peterson2021vesicle, vutukuri2020active}. However, coupling internal activity with vesicle shape to control different modes of locomotion, as occurs during amoeboid motion in living cells~\cite{ivvsic2026diversity}, has not yet been achieved in systems operating at microscopic to mesoscopic scales.

Capturing the physics of active motion, originally explored in the context of microswimming~\cite{Purcell1977, qiu2014swimming} and collective locomotion in animal groups~\cite{vicsek2012collective}, has motivated the development of synthetic Active Particle (AP) assemblies based on colloids~\cite{bishop2023active} or robotic swarms~\cite{bray2023recent}. At the single hard AP level, externally imposed cues have been successfully used to tune transport~\cite{ebbens2018catalytic, liebchen2018viscotaxis, bialus2025} or cause rectification, including ratchet-like~\cite{reichhardt2017ratchet} and oscillatory motion~\cite{bacot2019, Safara2022}. In AP assemblies, emergent transport was found to rely on many-body interactions~\cite{mandal2024} with control parameters such as AP packing fraction~\cite{morin2017diffusion}, chirality~\cite{tan2022odd, vega2022diffusive, Kiechl2026}, activity~\cite{deblais2018boundaries, boudet2021collections}, and even interface deformation ~\cite{ hooshanginejad2024}. However, an individual soft AP that can autonomously generate and modulate its locomotion as a synthetic analogue of a single motile cell is still missing, even at mesoscopic to macroscopic scales~\cite{chen2017rotational, lee2023complex, deblais2018boundaries, boudet2021collections, le2022encapsulated}. Consequently, so far, one could not explore shape deformability as an internal control parameter that enables programmable transport regimes of a single soft AP, without invoking collective effects or active phase transitions.

\begin{figure}[t!]
    \centering
    \includegraphics[width=\columnwidth]{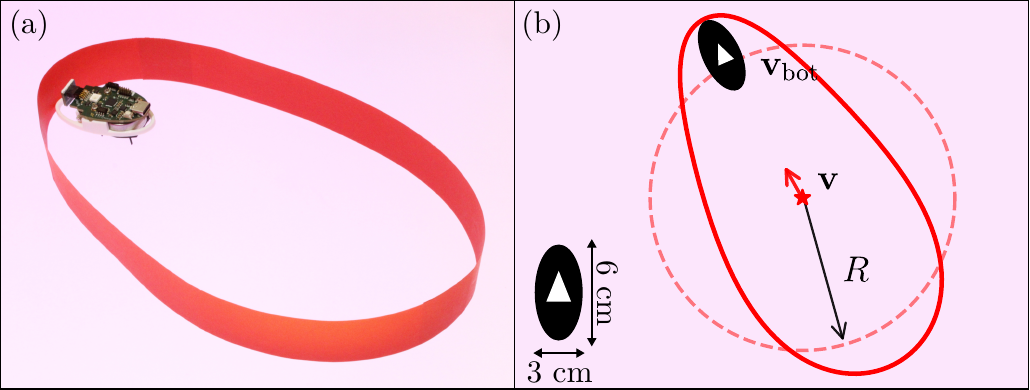}
    \caption{(a) Experimental setup consisting of a GRASPion confined to a soft membrane (in red). (b) Deformation and translation of a 2D membrane of radius $R$ with velocity $\mathbf{v}$ due to bot action $\mathbf{v}_{\mathrm{bot}}$, as implemented in simulations.} 
    \label{fig:Arena}
\end{figure}

Here, we demonstrate such behavior in a minimal soft robotic cell (Fig.~\ref{fig:Arena}). Our experimental setup consists of a deformable membrane enclosing a GRASPion, a centimeter-sized, programmable, elliptical robotic AP~\cite{novkoski_2025_graspion, noirhomme2025brainbots, mammadli2026physics}. To form a deformable active `cell', the bot is confined within a paper ring ($\qty{80}{\gram.\meter^{-2}}$, $\qty{94.8}{\micro\meter}$ thick) of height $\qty{3}{\centi\meter}$ and radius $R \in [\qty{4}, \qty{13}]$ \unit{\centi\meter}. Increasing $R$ effectively modulates the cell softness as the energy required for its deformation reduces as $R^{-3}$ (Supplementary Information~\cite{suppmat}). \nocite{singer2008lectures,bickel2007note,galassi2018scientific}

To determine non-trivial effects of softness on cell motility, we program the bot to exhibit a ballistic motion at short time scales and diffusive motion at long time scales~\cite{noirhomme2025brainbots}. This behavior is generated using a stochastic propulsion protocol in which periods of bot translation alternate with rotations, the latter occurring about a point offset by $\qty{1}{\centi\meter}$ from the bot center along the minor axis. The duration of each phase is sampled from a uniform distribution $\mathcal{U}[\qty{0.4}{\second},\qty{1.2}{\second}]$, with clockwise and anticlockwise rotations equally likely. Consequently, the bot behaves as a persistent AP with mean free translation velocity $v_{\mathrm{bot}} = \qty{15 \pm 0.2}{\centi\meter.\second^{-1}}$, persistence time $\tau_{\mathrm{bot}} = \qty{0.8 \pm 0.1}{\second}$, and long-time diffusion constant $D_{\mathrm{b}} = \qty{12.8 \pm 0.1}{\centi\meter^2.\second^{-1}}$.  

\begin{figure*}[t!]
    \centering
    \includegraphics[width=\textwidth]{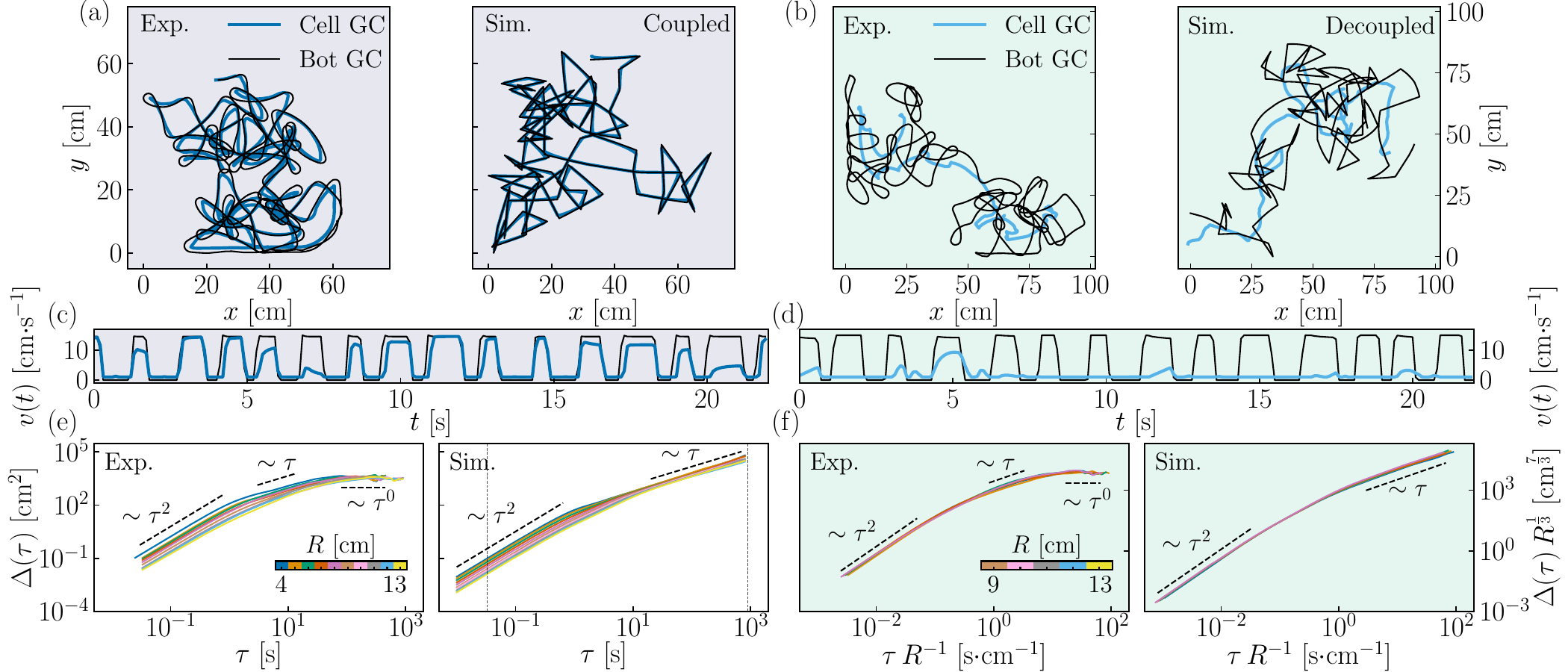}
    \caption{Representative trajectories of the geometric center (GC) of the cell and the bot, corresponding to (a) coupled dynamics (shaded purple: membrane of radius $R=\qty{4}{\centi\meter}$), and (b) decoupled dynamics (shaded green: membrane of radius $R=\qty{12}{\centi\meter}$), respectively. Panels (c) and (d) compare the time evolution of the absolute value of the velocity of the GCs of the bot and the cell in the coupled and decoupled states, respectively. Panel (e) shows the mean squared displacement (MSD) of the cell GC against time lag $\tau$ for different membrane radii $R$, calculated from experiments and simulations. Panel (f) scales the MSD by $R^{1/3}$ and $\tau$ by $R^{-1}$, leading to a collapse of the ballistic and the crossover regimes in the decoupled dynamical state. (Note: The dashed vertical lines in panel (e) represent the range of time lags from the experiments.)}
    \label{fig:traj_MSD}
\end{figure*}

The cell is made to move on a horizontal \qty{1}{\meter}$\times$\qty{1}{\meter} plexiglass plate.  A top-mounted camera is used to record its motion at $\qty{30}{frames/\second}$. Image analysis (\cite{suppmat}, Sec.~S1), is then used to extract the position of the bot, the membrane, and its instantaneous shape for further processing.
 
Our experiments are complemented by simulations in which the bot and the membrane are modeled as two-dimensional interacting entities evolving via overdamped dynamics in an open domain (\cite{suppmat}, Sec.~S2). The bot shape and driving protocol are parametrized as in experiments, except that rotations occur about the bot center, resulting in sharper turns (Fig.~\ref{fig:traj_MSD}a,b). 

The time evolution of the bot state is given by
\begin{equation}\label{EQ:bot_dynamics}
   \bm{\mathfrak{F}}_\mathrm{dr}+\bm{\mathfrak{F}}_\mathrm{int}=\gamma_\mathrm{bot}^\mathrm{T}\frac{\dd \bm{r}_\mathrm{bot}}{\dd t}, \quad \mathfrak{T}_\mathrm{dr}+\mathfrak{T}_\mathrm{int}=\gamma_\mathrm{bot}^\mathrm{R}\frac{\dd \vartheta_\mathrm{bot}}{\dd t},
\end{equation}
where $\bm{\mathfrak{F}}_\mathrm{dr}$ and $\mathfrak{T}_\mathrm{dr}$ denote the instantaneous driving force and torque of the bot, and $\bm{\mathfrak{F}}_\mathrm{int}$ and $\mathfrak{T}_\mathrm{int}$ are the corresponding membrane--bot interaction terms. Parameters $\gamma_\mathrm{bot}^T$ and $\gamma_\mathrm{bot}^R$ are the translational and rotational friction coefficients, respectively. During translation $\bm{\mathfrak{F}}_\mathrm{dr}=F_0\bm{\hat e}(\vartheta_\mathrm{bot})$ and $\mathfrak{T}_\mathrm{dr}=0$, while during rotation $\bm{\mathfrak{F}}_\mathrm{dr}=0$ and $\mathfrak{T}_\mathrm{dr}=\pm T_0$, where $F_0$ and $T_0$ are constant force and torque magnitudes, and $\bm{\hat e}(\vartheta_\mathrm{bot})$ is the unit vector along the bot orientation.

The membrane is modeled as a closed, discretized, non-intersecting curve, and its motion is described by 
\begin{equation}\label{EQ:band_overdamped}
\bm{\mathfrak{f}}_\mathrm{b}\left(\xi\right)+\bm{\mathfrak{f}}_\mathrm{st}\left(\xi\right)+\bm{\mathfrak{f}}_\mathrm{int}\left(\xi\right)=\bm{v}\left(\xi\right)\gamma_\mathrm{mem} R,
\end{equation}
where $\xi\in\left[0,2\pi\right)$ is parameterizing the band.
Here, $\bm{\mathfrak{f}}_\mathrm{b}$, $\bm{\mathfrak{f}}_\mathrm{st}$, and $\bm{\mathfrak{f}}_\mathrm{int}$, respectively, denote the bending, surface tension, and bot--membrane forces acting on a membrane segment $\dd \xi$. Velocity of the membrane segment is denoted by $\bm{v}(\xi)$, $\gamma_\mathrm{mem}$ is the friction coefficient per unit length, and the length of the segment is approximately $R\dd \xi$ (due to high surface tension, the total membrane length $2\pi R$ remains fixed within $\pm 0.1\%$). In the model, once the membrane loses contact with the bot, it relaxes toward its circular equilibrium shape via dissipation-mediated dynamics (\cite{suppmat}, Sec.~S2), in contrast with experiment, where such relaxation is negligible due to static friction, which is not included in the simulations.

\begin{figure*}[t!]
    \centering
    \includegraphics[width=\textwidth]{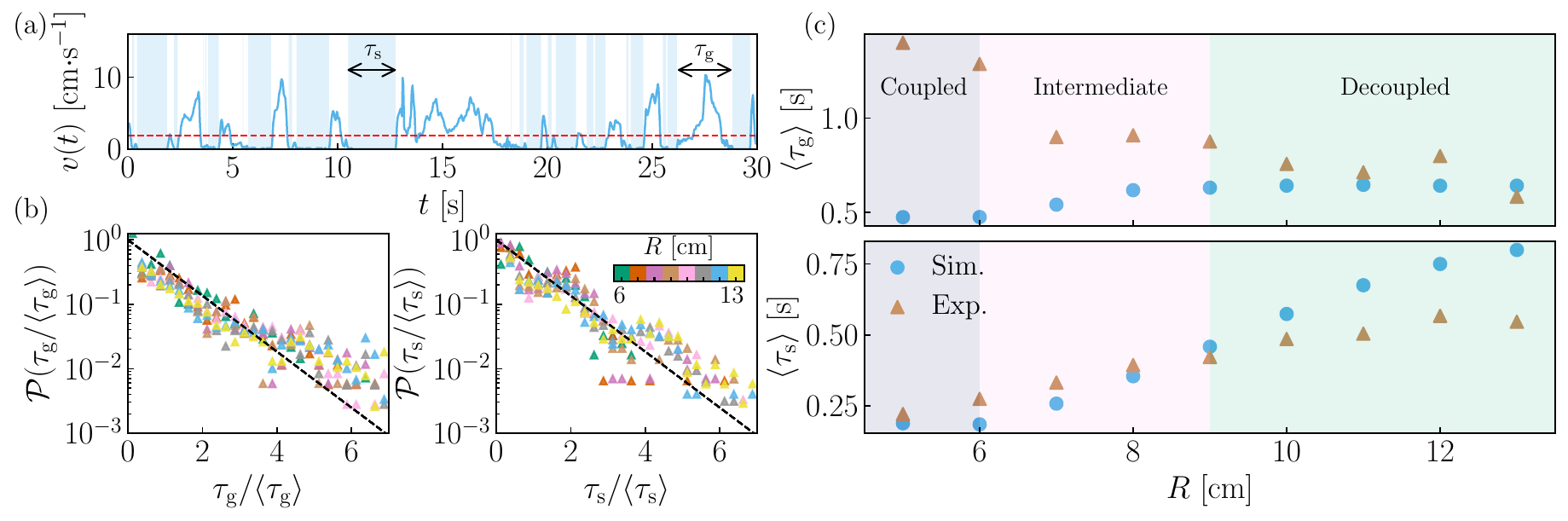}
    \caption{(a) Velocity time series of the geometric center (GC) of a cell of radius $R=\qty{12}{\centi\meter}$, with the shaded and unshaded areas demarcating `stop' and `go' durations as random variables $\tau_\mathrm{s}$ and $\tau_\mathrm{g}$, respectively. The dashed red line represents the time-averaged velocity of the cell GC. (b) Normalized PDFs $\mathcal{P}(\cdot)$ of go times (left) and stop times (right) for cells of different sizes as measured in experiments ($\qty{6} \leq R \leq\qty{13}{\centi\meter}$, each shown by a different color). The points represent the heights of the bins, while the dashed line is an exponential with an exponent of $-1$. See~\cite{suppmat} for details of the analysis. (c) Variation in average go time $\langle\tau_\mathrm{g}\rangle$ (top) and average stop time $\langle\tau_\mathrm{s}\rangle$ (bottom) across membrane sizes represented by radius $R$ for experiments (orange $\triangle$), and simulations (cyan $\circ$). The shading refers to the two phenomenological regimes of the cell: coupled (purple) and decoupled (green), and the intermediate crossover regime (pink).}
    \label{fig:tau_stats}
\end{figure*}

In both experiments and simulations, when the membrane size is comparable to that of the bot ($R=[4,5]$\unit{\centi\meter}), we observe coupled behavior, where the geometric centers (GCs) of the bot and the cell always move together (purple background in Fig.~\ref{fig:traj_MSD} and subsequent figures). Hence, the bot rotations and translations map directly to the stop and go phases of the cell, respectively (Fig.~\ref{fig:traj_MSD}c). Upon increasing membrane radius, there is a continuous crossover to a decoupled regime (green background in Fig.~\ref{fig:traj_MSD} and subsequent figures), where the GCs of the bot and the cell are effectively dissociated (Fig.~\ref{fig:traj_MSD}b). Therefore, the still prevailing stop-and-go dynamics of the cell no longer reflect the bot's driving protocol (Fig.~\ref{fig:traj_MSD}d).   

The stop-and-go dynamics underpins distinct transport properties of the synthetic cell, characterized by the mean-squared displacement $\Delta(\tau)$ (Fig.~\ref{fig:traj_MSD}e). The cell exhibits ballistic scaling at short time lags $\tau$, and a crossover to diffusive scaling at long time lags, with additional finite-size effects observed for experimental data. Upon increasing $R$, the increase in effective membrane softness leads to longer crossover regimes in $\Delta$ (see Fig.~\ref{fig:traj_MSD}e). 
In the decoupled regime, the effect of softness is, furthermore, evidenced by a collapse of the curves for the ballistic and crossover regimes with $\Delta$ scaling as $R^{1/3}$ and the time lag with $R^{-1}$ (Fig.~\ref{fig:traj_MSD}f). The same scaling applies to both experiments and simulations, thus validating the modeling approach. We note, however, that this scaling is not universal and depends on the bot's driving protocol and the membrane bending stiffness. 

Further insight into the transport properties of the robotic cell can be obtained by isolating the stop and go phases from trajectories 
(Fig.~\ref{fig:tau_stats}a) and extracting their associated distributions. In experiments, both distributions are exponential (Fig.~\ref{fig:tau_stats}b). Nonetheless, the go-phase statistics in small cells reflect uncertainty in experimental image tracking and the subsequent thresholding of the velocity signal (\cite{suppmat}, Sec.~S3). In simulations (Fig. S2), owing to the finite relaxation time of the membrane, the go-phase distributions are affected at short time scales. As a result, the mean go time $\langle\tau_\mathrm{g}\rangle$ in experimental and simulated trajectories 
(Fig.~\ref{fig:tau_stats}c, top) show good agreement only in the decoupled regime. 

The mean stop time $\langle\tau_\mathrm{s}\rangle$ 
(Fig.~\ref{fig:tau_stats}c, bottom), by contrast, shows good overall agreement between 
experiments and simulations, with noticeable deviations only at large membrane sizes. These 
are likely due to static friction, which, in experiments, allows parts of the cell to deform 
without significantly displacing its center of mass. This effect is absent in simulations, where the cell 
always moves when the bot and membrane interact, resulting in systematically longer go times 
compared to experiments.

Turning to the dependence on membrane size, both in experiments and simulations, $\langle\tau_s\rangle$ follows a monotonically increasing trend (Fig.~\ref{fig:tau_stats}c, bottom), simply reflecting the fact that the time the bot spends moving inside the cell without directly interacting with the membrane grows with cell radius. Meanwhile, the mean go time $\langle \tau_\mathrm{g} \rangle$ captures the influence of membrane softness (Fig.~\ref{fig:tau_stats}c, top). As softness increases beyond the coupled state, growing membrane deformations progressively constrict the bot orientation, promoting longer unidirectional bot excursions, particularly in the intermediate regime. Beyond a certain softness threshold in the decoupled regime, however, even large cell deformations can no longer constrict the bot, and $\langle \tau_\mathrm{g} \rangle$ saturates.


The exponential distributions of stop and go durations (Fig.~\ref{fig:tau_stats}b) naturally motivate modeling the cell as a two-state Markovian run-and-tumble particle. Accordingly, the propulsion velocity of the run phase is drawn from a random process, and the propulsion direction is fully randomized at each tumble, an approximation expected to break down in the coupled regime. With negligible thermal diffusion, an analytic form of $\Delta(\tau)$ is then derived from the velocity autocorrelation (\cite{suppmat}, Sec.~S4), yielding
\begin{align}
\Delta(\tau) &= 2\hat{v}_0^2 \Bigl(\tau\,\langle\tau_\mathrm{g}\rangle - \langle\tau_\mathrm{g}\rangle^{\,2}\left(1 - e^{\tau/\langle\tau_\mathrm{g}\rangle}\right)\Bigr),
\label{eq:MSD}
\end{align}
where
\begin{equation}
\hat{v}_0 = \sqrt{\bar{v}^2 + \sigma_v^2},
\label{eq:vzero}
\end{equation}
and $\bar{v}$ and $\sigma_v^2$ are the mean and the variance of the full velocity distribution, including the zero-velocity stop events. At short time scales ($\tau < \tau^\ast$, $\tau^\ast$ being the crossover time), the MSD is a quadratic function in time lag, $\Delta \left(\tau\right) \simeq (\hat{v}_0\, \tau)^2$. In the diffusive regime, at long time scales ($\tau\gg\tau^\ast$), $\Delta \left(\tau \right) \simeq \hat{D}\,\tau$, with 
$\hat{D}=2\, \hat{v}_0^{2}\, \langle\tau_\mathrm{g}\rangle$
being the diffusion coefficient of the cell.

\begin{figure}[t!]
    \centering
    \includegraphics[width=\linewidth]{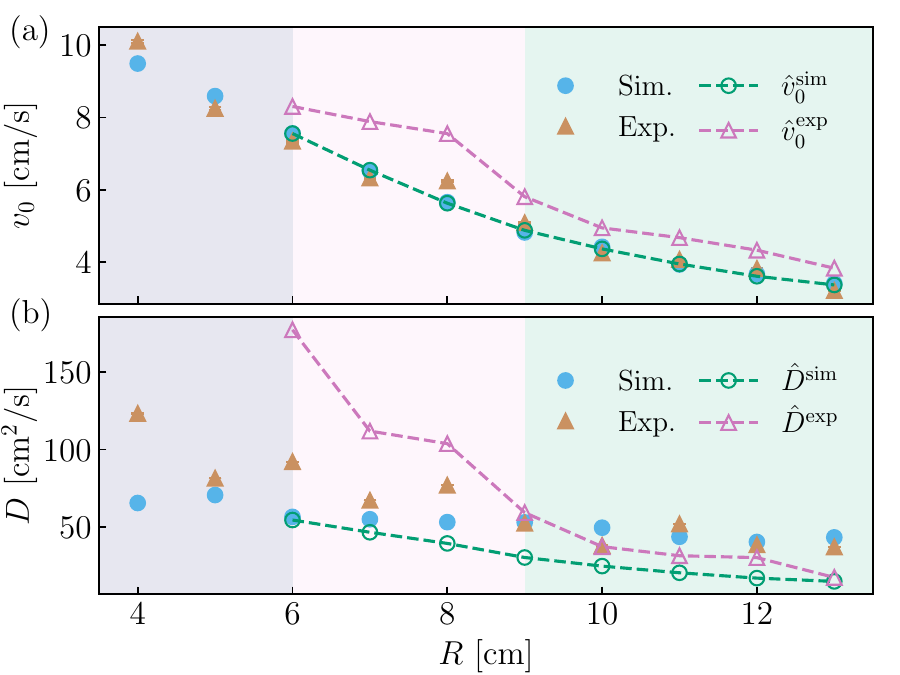}
    \caption{Comparison of theoretical predictions with experiments and simulations. (a) Ballistic velocity extracted by fitting the MSD at short times, compared to theoretical predictions via Eq.~\ref{eq:vzero}, using the first and second moments of the velocity distribution from experiments ($\hat{v}_0^\mathrm{exp}$) and simulations ($\hat{v}_0^\mathrm{sim}$). (b) Diffusion coefficient extracted by fitting the MSD at long times, compared to theoretical predictions, using statistics from experiments ($\hat{D}^\mathrm{exp}$) and simulations ($\hat{D}^\mathrm{sim}$).}
    \label{fig:msd_coeffs}
\end{figure}

To compare this model with simulations and experiments, we first determine the ballistic velocity and the diffusion coefficients from the data by regression of the  $\Delta (\tau)$ (Fig.~\ref{fig:traj_MSD}e) in short and long time limits, respectively. We find that the ballistic velocity decreases with increasing cell size (Fig.~\ref{fig:msd_coeffs}a), as a larger fraction of the active input is directed toward membrane deformation and frictional dissipation, the latter being proportional to $R$. The reduction in mobility with size also translates into the reduction of the diffusion coefficient with increasing membrane size (Fig.~\ref{fig:msd_coeffs}b).

The theoretical model well reproduces these trends. The ballistic velocity $\hat{v}_0$, calculated from experimental and simulation velocity distributions, is consistent with the expected short-time ballistic scaling of run-and-tumble dynamics~\cite{dinelli2024fluctuating}(Fig.~\ref{fig:msd_coeffs}a). The theoretical diffusion coefficient $\hat{D}^\mathrm{sim}$ also shows fair overall agreement with fits of $\Delta(\tau)$ from simulations, with a systematic offset emerging at large membrane sizes due to an underestimation of velocity correlations within the model (Fig.~\ref{fig:msd_coeffs}b). The experimental counterpart $\hat{D}^\mathrm{exp}$, by contrast, deviates in the intermediate regime due to an overestimation of the mean go-time during the analysis procedure. Despite these deviations, the theory captures well the overall transport properties of the robotic cell, validating the run-and-tumble framework as an effective and minimal description of its locomotion statistics.

These findings reflect a broader result: the synthetic robotic cell introduced here exhibits transport properties that are highly tunable through effective membrane softness, which acts as a single control parameter governing the transition between distinct dynamical regimes, from coupled to decoupled, and from ballistic to diffusive transport. This tunability fulfills a long-standing goal in active matter by realizing a minimal, programmable soft AP with continuously controllable complex transport properties. In turn, this offers a tractable platform that captures key physical properties of living motile cells, without invoking the full complexity of their biological counterparts.

The programmability of our robotic cell, therefore, opens the door to mimicking different locomotion strategies~\cite{mammadli2026physics}, with softness as an independent tuning parameter. This is particularly relevant for studies of transport through complex environments, where different propulsion strategies are expected to reflect differently in the capacity of the cell to navigate obstacles and confinement. Scaling up to assemblies of many such soft robotic cells would further allow studies of cooperative effects directly analogous to those in living tissues. Finally, extending the model toward active control of cell shape itself~\cite{Jones2021}, as living cells do during amoeboid motion, would introduce a feedback mechanism between locomotion and deformation, potentially giving rise to spontaneous symmetry breaking, dynamic shifts between transport regimes, and oscillatory or chaotic states, ultimately enabling emergent adaptive locomotion strategies in a fully synthetic robotic system. The soft robotic active particle presented herein is therefore a crucial milestone toward truly lifelike artificial cells.


\textbf{Acknowledgments:} This work was supported by FNRS through the grant PDR T.0251.20, the Deutsche Forschungsgemeinschaft (DFG, German Research Foundation) through Project-ID 416229255 - SFB 1411 Design of Particulate Products (subproject D01), and the Croatian Science Foundation with the grant IP-2025-02-3976. F.N. thanks the Alexander von Humboldt Foundation for a postdoctoral fellowship. Special thanks to M.~Mélard for his technical support in developing the GRASPion. For the computational work, we acknowledge the support of SRCE, Croatia, and NHR@FAU, Germany.

\bibliographystyle{apsrev4-2} 
\bibliography{bibliography} 

@article{Jones2021,
  author  = {Trevor J. Jones and Etienne Jambon-Puillet and Joel Marthelot and P.-T. Brun},
  title   = {Bubble casting soft robotics},
  journal = {Nature},
  volume  = {599},
  number  = {7884},
  pages   = {229--233},
  year    = {2021},
  doi     = {10.1038/s41586-021-04029-6}
}

@article{mandal2024,
  title     = {Robustness of traveling states in generic nonreciprocal mixtures},
  author    = {Mandal, Rituparno and Salazar Jaramillo, Santiago and Sollich, Peter},
  journal   = {Phys. Rev. E},
  volume    = {109},
  number    = {6},
  pages     = {L062602},
  year      = {2024},
  doi       = {10.1103/PhysRevE.109.L062602},
  publisher = {American Physical Society}
}

@article{bialus2025,
  title     = {Enhancement and suppression of active particle movement due to membrane deformations},
  author    = {Bialus, Adam Hitin and Rallabandi, Bhargav and Oppenheimer, Naomi},
  journal   = {J. Fluid Mech.},
  volume    = {1024},
  pages     = {A42},
  year      = {2025},
  doi       = {10.1017/jfm.2025.10836},
  publisher = {Cambridge University Press}
}

@article{Safara2022,
  author  = {Francisco M. R. Safara and Hygor P. M. Melo and Margarida M. Telo da Gama and Nuno A. M. Ara{\'u}jo},
  title   = {Model for active particles confined in a two-state micropattern},
  journal = {Soft Matter},
  volume  = {18},
  pages   = {5699--5705},
  year    = {2022},
  doi     = {10.1039/D2SM00616B}
}

@article{bacot2019,
  title     = {Multistable free states of an active particle from a coherent memory dynamics},
  author    = {Bacot, Vincent and Perrard, St\'ephane and Labousse, Matthieu and Couder, Yves and Fort, Emmanuel},
  journal   = {Phys. Rev. Lett.},
  volume    = {122},
  number    = {10},
  pages     = {104303},
  year      = {2019},
  doi       = {10.1103/PhysRevLett.122.104303},
  publisher = {American Physical Society}
}

@article{hooshanginejad2024,
  title     = {Interactions and pattern formation in a macroscopic magnetocapillary {SALR} system of mermaid cereal},
  author    = {Hooshanginejad, Alireza and Barotta, Jack-William and Spradlin, Victoria and Pucci, Giuseppe and Hunt, Robert and Harris, Daniel M.},
  journal   = {Nat. Commun.},
  volume    = {15},
  number    = {1},
  pages     = {5466},
  year      = {2024},
  doi       = {10.1038/s41467-024-49754-4},
  publisher = {Nature Publishing Group}
}

@article{Kiechl2026,
  author    = {Thomas Kiechl and Amy Altshuler and Anton L{\"u}ders and Yael Roichman and Thomas Franosch},
  title     = {Free chiral self-propelled robots compared to active Brownian circle swimmers},
  journal   = {Physical Review E},
  year      = {2026},
  volume     = {113},
  number     = {4},
  pages      = {045409},
  publisher  = {American Physical Society},
  month      = apr,
  doi        = {},
}

@article{dinelli2024fluctuating,
  title={Fluctuating hydrodynamics of active particles interacting via taxis and quorum sensing: static and dynamics},
  author={Dinelli, Alberto and O’Byrne, J{\'e}r{\'e}my and Tailleur, Julien},
  journal={J. Phys. A: Math. Theor.},
  volume={57},
  number={39},
  pages={395002},
  year={2024},
  publisher={IOP Publishing}
}

@article{novkoski_2025_graspion,
  title={GRASPion: An open-source, programmable brainbot for active matter research},
  author={Novkoski, Filip and M{\'e}lard, M{\'e}d{\'e}ric and Delens, Megan and W{\'e}ry, Fanny and Noirhomme, Martial and Pande, Jayant and Maier, Andreas and Smith, Ana-Suncana and Vandewalle, Nicolas},
  journal={Rev. Sci. Instru.},
  volume={97},
  number={1},
  year={2026},
  publisher={AIP Publishing}
}

@article{bickel2007note,
  title={A note on confined diffusion},
  author={Bickel, Thomas},
  journal={Phys. A: Stat. Mech. Appl.},
  volume={377},
  number={1},
  pages={24--32},
  year={2007},
  publisher={Elsevier}
}

@article{bechinger2016active,
  title={Active particles in complex and crowded environments},
  author={Bechinger, Clemens and Di Leonardo, Roberto and L{\"o}wen, Hartmut and Reichhardt, Charles and Volpe, Giorgio and Volpe, Giovanni},
  journal={Rev. Mod. Phys.},
  volume={88},
  number={4},
  pages={045006},
  year={2016},
  publisher={APS}
}

@article{vicsek2012collective,
  title={Collective motion},
  author={Vicsek, Tam{\'a}s and Zafeiris, Anna},
  journal={Phys. Rep.},
  volume={517},
  number={3-4},
  pages={71--140},
  year={2012},
  publisher={Elsevier}
}

@article{cates2015motility,
  title={Motility-induced phase separation},
  author={Cates, Michael E and Tailleur, Julien},
  journal={Annu. Rev. Condens. Matter Phys.},
  volume={6},
  number={1},
  pages={219--244},
  year={2015},
  publisher={Annual Reviews}
}

@article{marchetti2013hydrodynamics,
  title={Hydrodynamics of soft active matter},
  author={Marchetti, M Cristina and Joanny, Jean-Fran{\c{c}}ois and Ramaswamy, Sriram and Liverpool, Tanniemola B and Prost, Jacques and Rao, Madan and Simha, R Aditi},
  journal={Rev. Mod. Phys.},
  volume={85},
  number={3},
  pages={1143--1189},
  year={2013},
  publisher={APS}
}

@article{peterson2021vesicle,
  title={Vesicle shape transformations driven by confined active filaments},
  author={Peterson, Matthew SE and Baskaran, Aparna and Hagan, Michael F},
  journal={Nat. Commun.},
  volume={12},
  number={1},
  pages={7247},
  year={2021},
  publisher={Nature Publishing Group UK London}
}

@article{lee2023complex,
  title={Complex motion of steerable vesicular robots filled with active colloidal rods},
  author={Lee, Sophie Y and Sch{\"o}nh{\"o}fer, Philipp WA and Glotzer, Sharon C},
  journal={Sci. Rep.},
  volume={13},
  number={1},
  pages={22773},
  year={2023},
  publisher={Nature Publishing Group UK London}
}

@article{gandikota2023rectification,
  title={Rectification of confined soft vesicles containing active particles},
  author={Gandikota, MC and Cacciuto, A},
  journal={Soft Matter},
  volume={19},
  number={2},
  pages={315--320},
  year={2023},
  publisher={Royal Society of Chemistry}
}

@article{boudet2021collections,
  title={From collections of independent, mindless robots to flexible, mobile, and directional superstructures},
  author={Boudet, Jean-Fran{\c{c}}ois and Lintuvuori, J and Lacouture, Claire and Barois, Thomas and Deblais, Antoine and Xie, Kaili and Cassagnere, S{\'e}bastien and Tregon, Bernard and Br{\"u}ckner, David B and Baret, Jean-Christophe and others},
  journal={Sci. Robotics},
  volume={6},
  number={56},
  pages={eabd0272},
  year={2021},
  publisher={American Association for the Advancement of Science}
}

@article{deblais2018boundaries,
  title={Boundaries control collective dynamics of inertial self-propelled robots},
  author={Deblais, Antoine and Barois, Thomas and Guerin, T and Delville, Pierre-Henri and Vaudaine, R{\'e}mi and Lintuvuori, Juho S and Boudet, Jean-Fran{\c{c}}ois and Baret, Jean-Christophe and Kellay, H},
  journal={Phys. Rev. Lett.},
  volume={120},
  number={18},
  pages={188002},
  year={2018},
  publisher={APS}
}

@article{paoluzzi2016shape,
  title={Shape and displacement fluctuations in soft vesicles filled by active particles},
  author={Paoluzzi, Matteo and Di Leonardo, Roberto and Marchetti, M Cristina and Angelani, Luca},
  journal={Sci. Rep.},
  volume={6},
  number={1},
  pages={34146},
  year={2016},
  publisher={Nature Publishing Group UK London}
}

@article{chen2017rotational,
  title={Rotational diffusion of soft vesicles filled by chiral active particles},
  author={Chen, Jiamin and Hua, Yunfeng and Jiang, Yangwei and Zhou, Xiaolin and Zhang, Linxi},
  journal={Sci. Rep.},
  volume={7},
  number={1},
  pages={15006},
  year={2017},
  publisher={Nature Publishing Group UK London}
}

@article{uplap2023design,
  title={Design principles for transporting vesicles with enclosed active particles},
  author={Uplap, Sarvesh and Hagan, Michael F and Baskaran, Aparna},
  journal={Europhys. Lett.},
  volume={143},
  number={6},
  pages={67001},
  year={2023},
  publisher={EDP Sciences, IOP Publishing and Societ{\`a} Italiana di Fisica}
}

@article{noirhomme2025brainbots,
  title = {Brainbots as smart autonomous active particles with programmable motion},
  author = {Noirhomme, Martial and Mammadli, Isa and Vanesse, Nathan and Pande, Jayant and Smith, Ana-Suncana and Vandewalle, Nicolas},
  journal = {Phys. Rev. Appl.},
  volume = {23},
  issue = {6},
  pages = {064008},
  numpages = {10},
  year = {2025},
  month = {Jun},
  publisher = {American Physical Society},
  doi = {10.1103/pv3d-7s54},
  url = {https://link.aps.org/doi/10.1103/pv3d-7s54}
}

@misc{mammadli2026physics,
      title={Physics-informed digital twin and onboard control of a brainbot for intelligent active matter}, 
      author={Isa Mammadli and Prajol Shrestha and Jayant Pande and Filip Novkoski and Siddhant Mohapatra and Martial Noirhomme and Andreas Maier and Nicolas Vandewalle and Ana-Suncana Smith},
      year={2026},
      eprint={2511.00384},
      archivePrefix={arXiv},
      primaryClass={cond-mat.soft},
      url={https://arxiv.org/abs/2511.00384}, 
}

@article{Purcell1977,
  author = {Purcell, E. M.},
  title = {Life at low Reynolds number},
  journal = {Amer. J. Phys.},
  volume = {45},
  number = {1},
  pages = {3--11},
  year = {1977},
  doi = {10.1119/1.10903}
}

@article{qiu2014swimming,
  title={Swimming by reciprocal motion at low Reynolds number},
  author={Qiu, Tian and Lee, Tung-Chun and Mark, Andrew G and Morozov, Konstantin I and M{\"u}nster, Raphael and Mierka, Otto and Turek, Stefan and Leshansky, Alexander M and Fischer, Peer},
  journal={Nat. Commun.},
  volume={5},
  number={1},
  pages={5119},
  year={2014},
  publisher={Nature Publishing Group UK London}
}

@article{bishop2023active,
  title={Active colloids as models, materials, and machines},
  author={Bishop, Kyle JM and Biswal, Sibani Lisa and Bharti, Bhuvnesh},
  journal={Annu. Rev. Chem. Biomol. Eng.},
  volume={14},
  pages={1--30},
  year={2023},
  publisher={Annual Reviews}
}

@article{bray2023recent,
  title={Recent developments in self-assembling multi-robot systems},
  author={Bray, Edward and Gro{\ss}, Roderich},
  journal={Curr. Robo. Rep.},
  volume={4},
  number={4},
  pages={101--116},
  year={2023},
  publisher={Springer}
}

@article{ebbens2018catalytic,
  title={Catalytic Janus colloids: controlling trajectories of chemical microswimmers},
  author={Ebbens, Stephen J and Gregory, David Alexander},
  journal={Acc. Chem. Res.},
  volume={51},
  number={9},
  pages={1931--1939},
  year={2018},
  publisher={ACS Publications}
}

@article{liebchen2018viscotaxis,
  title={Viscotaxis: Microswimmer navigation in viscosity gradients},
  author={Liebchen, Benno and Monderkamp, Paul and Ten Hagen, Borge and L{\"o}wen, Hartmut},
  journal={Phys. Rev. Lett.},
  volume={120},
  number={20},
  pages={208002},
  year={2018},
  publisher={APS}
}

@article{reichhardt2017ratchet,
  title={Ratchet effects in active matter systems},
  author={Reichhardt, CJ Olson and Reichhardt, Charles},
  journal={Annu. Rev. Condens. Matter Phys.},
  volume={8},
  pages={51--75},
  year={2017},
  publisher={Annual Reviews}
}

@article{morin2017diffusion,
  title={Diffusion, subdiffusion, and localization of active colloids in random post lattices},
  author={Morin, Alexandre and Lopes Cardozo, David and Chikkadi, Vijayakumar and Bartolo, Denis},
  journal={Phys. Rev. E},
  volume={96},
  number={4},
  pages={042611},
  year={2017},
  publisher={APS}
}

@article{vutukuri2020active,
  title={Active particles induce large shape deformations in giant lipid vesicles},
  author={Vutukuri, Hanumantha Rao and Hoore, Masoud and Abaurrea-Velasco, Clara and Van Buren, Lennard and Dutto, Alessandro and Auth, Thorsten and Fedosov, Dmitry A and Gompper, Gerhard and Vermant, Jan},
  journal={Nature},
  volume={586},
  number={7827},
  pages={52--56},
  year={2020},
  publisher={Nature Publishing Group UK London}
}

@article{le2022encapsulated,
  title={Encapsulated bacteria deform lipid vesicles into flagellated swimmers},
  author={Le Nagard, Lucas and Brown, Aidan T and Dawson, Angela and Martinez, Vincent A and Poon, Wilson CK and Staykova, Margarita},
  journal={Proc. Nat. Acad. Sci.},
  volume={119},
  number={34},
  pages={e2206096119},
  year={2022},
  publisher={National Academy of Sciences}
}

@article{vega2022diffusive,
  title={Diffusive regimes in a two-dimensional chiral fluid},
  author={Vega Reyes, Francisco and L{\'o}pez-Casta{\~n}o, Miguel A and Rodr{\'\i}guez-Rivas, {\'A}lvaro},
  journal={Commun. Phys.},
  volume={5},
  number={1},
  pages={256},
  year={2022},
  publisher={Nature Publishing Group UK London}
}

@article{tan2022odd,
  title={Odd dynamics of living chiral crystals},
  author={Tan, Tzer Han and Mietke, Alexander and Li, Junang and Chen, Yuchao and Higinbotham, Hugh and Foster, Peter J and Gokhale, Shreyas and Dunkel, J{\"o}rn and Fakhri, Nikta},
  journal={Nature},
  volume={607},
  number={7918},
  pages={287--293},
  year={2022},
  publisher={Nature Publishing Group UK London}
}

@misc{suppmat,
    note = "See Supplemental Material at [URL will be inserted by publisher] for details on I.~experimental data collection, II.~numerical model, III.~experimental and simulation data processing, and IV.~the derivation of Mean-Squared Displacement of the cell. Source files are provided on Zenodo doi:10.5281/zenodo.17511061."
}

@article{ivvsic2026diversity,
  title={Diversity in emergent cell locomotion from the coupling cytosolic and cortical Marangoni flows with reaction--diffusion dynamics},
  author={Iv{\v{s}}i{\'c}, Bla{\v{z}} and Vuli{\'c}, Dorijan and Weber, Igor and Nowakowski, Piotr and Smith, Ana-Sun{\v{c}}ana},
  journal={PLOS Comp. Biol.},
  volume={22},
  number={4},
  pages={e1014216},
  year={2026},
  publisher={Public Library of Science San Francisco, CA USA}
}

@misc{galassi2018scientific,
  title        = {GNU Scientific Library Reference Manual},
  author       = {Galassi, M. and Gough, B. and Rossi, F. and Booth, M. and Jungman, G. and Theiler, J. and Davies, J.},
  year         = {2018},
  edition      = {3rd},
  isbn         = {0954612078},
  url          = {https://www.gnu.org/software/gsl/}
}

@inproceedings{singer2008lectures,
  title={Lectures on elastic curves and rods},
  author={Singer, David A and Garay, Oscar J and Garcia-Rio, Eduardo and Vazquez-Lorenzo, Ramon},
  booktitle={AIP Conference Proceedings},
  number={1},
  pages={3},
  year={2008}
}

\pagebreak
\newpage

\setcounter{equation}{0}
\setcounter{figure}{0}
\setcounter{table}{0}
\setcounter{section}{0}

\renewcommand{\theequation}{S\arabic{equation}}
\renewcommand{\thefigure}{S\arabic{figure}}
\renewcommand{\thetable}{S\arabic{table}}
\renewcommand{\thesection}{S\arabic{section}}

\begin{center}
{\bf Supplementary Information:\\  Run and tumble dynamics of a soft robotic cell}\\
\end{center}

\section{Experimental data collection}

For each cell radius $R$, a single experiment was performed. Experiments lasted $900$ s for cells with radii between $4$ cm and $9$ cm, and $1800$ s for cells with radii between $10$ cm and $13$ cm. During each experiment, video recordings of the system were acquired (using the Basler acA3088-57um, 6 Mp) and subsequently processed to extract the relevant kinematic quantities. Videos were recorded at a frame rate of $30$ frames/s, except for the experiment with $R=4$ cm, which was recorded at $40$ frames/s.

The different components of the cell were tracked using contour detection implemented in Python using the OpenCV library. At each frame, all contours are detected and sorted by size. The contour with the largest surface area corresponds to the cell boundary, i.e. the membrane, and its geometric center provides the position of the cell within the arena. The position of the bot is then identified by further contour detection inside the contour of the membrane. The videos and datasets for the experiments can be found in [35].

\section{Numerical model for the robotic cell}
In this section, we present the model of motion of the system that we used for our numerical simulations. This section is divided into four parts, detailing the modeling of the membrane and the bot, followed by a simulation algorithm and the set of parameters used.

\subsection{Modeling the membrane}

We model the membrane as a closed, smooth, nonintersecting curve on a two-dimensional plane. We use vector parametrization
\begin{equation}
    \bm{\gamma}\left(\xi\right)=\left[ x\left(\xi\right), y\left(\xi\right) \right], \qquad \xi\in\left[0,2\pi\right],
\end{equation}
where $\xi$ is a parameter, and $x\left(\xi\right)$ and $y\left(\xi\right)$ are periodic functions. We denote the derivatives with respect to $\xi$ by dots, and the length element is
\begin{equation}
    \dd \ell=\dd \xi\sqrt{\dot{x}^2\left(\xi\right)+\dot{y}^2\left(\xi\right)}.
\end{equation}

\subsubsection{Energy of the membrane}
We define the energy of the membrane as a functional of its shape and configuration of bot $\bm{r}_\mathrm{bot}$ and $\vartheta_\mathrm{bot}$
\begin{equation}
    \mathfrak{E}\left(\left[\bm{\gamma}\left(\xi\right)\right], \bm{r}_\mathrm{bot},\vartheta_\mathrm{bot}\right)=\int_0^{2\pi}\dd \xi\left( \mathcal{E}_{\mathrm{b}}+\mathcal{E}_{\mathrm{st}}+\mathcal{E}_{\mathrm{int}}\right),
\end{equation}
where $\mathcal{E}_\mathrm{b}$, $\mathcal{E}_\mathrm{st}$, and $\mathcal{E}_\mathrm{int}$ denote, respectively, the densities of bending energy, surface tension energy, and interaction between the membrane and the bot.

The bending energy density is given by [36]
\begin{equation}\label{SI:EQ:Eel}
    \mathcal{E}_\mathrm{b}\left(\bm\gamma\left(\xi\right)\right)=k \kappa^2\left(\xi\right)\sqrt{\dot{x}^2\left(\xi\right)+\dot{y}^2\left(\xi\right)},
\end{equation}
where $k$ is a bending stiffness coefficient, and
\begin{equation}
    \kappa\left(\xi\right)=\frac{\dot{x}\left(\xi\right)\ddot{y}\left(\xi\right)-\dot{y}\left(\xi\right)\ddot{x}\left(\xi\right)}{\left[\dot{x}^2\left(\xi\right)+\dot{y}^2\left(\xi\right)\right]^{3/2}},
\end{equation}
is a local curvature of the membrane. The bending stiffness is given by the physical properties of the membrane
\begin{equation}
    k=\frac{1}{24}Y d^3 h,
\end{equation}
where $Y$ is the Young's modulus of the material, $d$ is the thickness, and $h$ denotes the height of the membrane. We note that the energy defined via Eq.~\eqref{SI:EQ:Eel} does not depend on the parametrization of the membrane.

We define the surface tension energy as
\begin{equation}\label{SI:EQ:Est}
    \mathcal{E}_\mathrm{st}\left(\bm\gamma\left(\xi\right)\right)=\alpha \frac{\ell_0}{2\pi}\left(2\pi\sqrt{\dot{x}^2\left(\xi\right)+\dot{y}^2\left(\xi\right)}-\ell_0\right)^2,
\end{equation}
where $\alpha$ is the surface tension coefficient and $\ell_0$ denotes the length of unstretched membrane. This term plays a twofold role. On the one hand, it ensures that the length of the membrane remains almost constant and equal to $\ell_0$, as observed in the experiment. On the other hand, it keeps the parametrization uniform by making
\begin{equation}\label{SI:EQ:dl}
    \sqrt{\dot{x}^2\left(\xi\right)+\dot{y}^2\left(\xi\right)}\approx \frac{\ell_0}{2\pi};
\end{equation}
\noindent We discuss the validity of this approximation below. The factor $\ell_0/\left(2\pi\right)$ has been included in Eq.~\eqref{SI:EQ:Est} to ensure the extensivity of the membrane energy.

Finally, for the energy of interaction between the membrane and the bot, we take
\begin{equation}\label{SI:EQ:Eint}
    \mathcal{E}_\mathrm{int}\left(\bm{\gamma}\left(\xi\right);\bm{r}_\mathrm{bot},\vartheta_\mathrm{bot}\right)=\frac{\ell_0}{2\pi}\mathfrak{V}\Big(\varrho\big(\bm{\gamma}\left(\xi\right);\bm{r}_\mathrm{bot},\vartheta_\mathrm{bot}\big)\Big),
\end{equation}
where $\varrho$ denotes the signed distance from the circumference of the elliptical bot, and $\mathfrak{V}$ is the potential of interaction. We define the signed distance $\varrho\big(\bm{r}_0;\bm{r}_\mathrm{bot},\vartheta_\mathrm{bot}\big)$ as the length of the shortest segment starting in $\bm{r}_0$ and ending at the circumference of the ellipse, multiplied by a factor of $-1$ if the point $\bm{r}_0$ is inside the ellipse. This ensures $\rho$ is a continuous function of $\bm{r}_0$ that is negative inside, zero on the surface of, and positive outside the ellipse. We take
\begin{equation}
    \mathfrak{V}\left(x\right)=A \exp\left(-x/\lambda\right),
\end{equation}
where $A$ is the amplitude and $\lambda$ is the length of the interaction, such that the interaction is negligible when a part of the membrane is far from the ellipse of the bot and very strong if it is inside. The prefactor $\ell_0/\left(2\pi\right)$ in Eq.~\ref{SI:EQ:Eint}, like in the previous case, makes the energy of interaction extensive.

\subsubsection{Deformation and the relation between cell softness and size}

The membrane in the current work equilibrates to a circular shape in the absence of external forces. We can parameterize the circle as
\begin{equation}
    x\left(\theta\right)=\frac{\ell_0}{2\pi}\cos \theta, \,\, y\left(\theta\right)=\frac{\ell_0}{2\pi}\sin \theta \qquad \theta\in\left[0,2\pi\right],
\end{equation}
where $\theta$ is the angle subtended by a membrane element to the center of the circle, and $\ell_0=2\pi R$ is the length of the membrane.

The bending energy associated with the membrane, following Eq.~\eqref{SI:EQ:Eel}, is
\begin{equation}
    \begin{aligned}\label{eq:b_energy}
    \mathcal{E}_{\mathrm{b}} &= k\int_0^{2\pi} \left(\left(\frac{\dot{x}\ddot{y} - \dot{y}\ddot{x}}{\left(\dot{x}^2 + \dot{y}^2 \right)^{3/2}} \right)^2 \sqrt{\dot{x}^2 + \dot{y}^2}\right)\, \dd\theta \\
    &= \frac{4k\pi^2}{\ell_0}.
    \end{aligned}
\end{equation}
In order to calculate how much energy is needed for a small deformation of the membrane, we assume that the circular shape is deformed slightly into an ellipse. The parametrization after this deformation is
\begin{equation}
    x'\left(\theta\right)=\left(\frac{\ell_0}{2\pi} + \epsilon \right)\cos \theta, \,\, y'\left(\theta\right)=\left(\frac{\ell_0}{2\pi} - a\epsilon\right)\sin \theta,
\end{equation}
where small positive $\epsilon$ measures the amplitude of deformation and $a$ is yet-to-be-determined parameter regulating ratio of deformations in $y$ and $x$ directions.

The length of the deformed membrane is:
\begin{multline}
    \ell_0^\prime = \int_0^{2\pi} \sqrt{\dot{x}'^2 + \dot{y}'^2}\, \dd\theta \\= \ell_0+(1-a)\pi\epsilon + \frac{(1+a)^2\pi^2\epsilon^2}{4\ell_0}+\mathrm{O}\left(\epsilon^3\right),
\end{multline}
which shows that if we take $a=1$, the length of the membrane is not changed by the deformation up to corrections of the order of $\epsilon^2$. Since the length of the membrane is fixed in our experiments, we take $a=1$. Following Eq.~\eqref{eq:b_energy}, the elastic energy after deformation is
\begin{equation}
\begin{aligned}
    \mathcal{E}_{\mathrm{b}}' &= k\int_0^{2\pi}\left(\left(\frac{\dot{x}'\ddot{y}' - \dot{y}'\ddot{x}'}{\left(\dot{x}'^2 + \dot{y}'^2 \right)^{3/2}} \right)^2 \sqrt{\dot{x}'^2 + \dot{y}'^2}\right)\, \dd\theta \\
    &= \mathcal{E}_{\mathrm{b}}+\frac{68k\pi^4 \epsilon^2}{\ell_0^3} + \mathrm{O}\left(\epsilon^3\right).
\end{aligned}
\end{equation}
Therefore, the energy associated with a small deformation scales as $\ell_0^{-3}$, i.e., $R^{-3}$, which explains why larger membranes are effectively softer and easier to deform.

\subsubsection{Time evolution of membrane motion}

The density of the forces acting on the membrane is calculated using the functional derivative as
\begin{equation}
    \bm{\mathfrak{F}}=-\left[\frac{\delta\mathfrak{E}}{\delta x\left(\xi\right)},\frac{\delta\mathfrak{E}}{\delta y\left(\xi\right)}\right]=\int_0^{2\pi}\dd\xi\left(\bm{\mathfrak{f}}_\mathrm{b}+\bm{\mathfrak{f}}_\mathrm{st}+\bm{\mathfrak{f}}_\mathrm{int}\right),
\end{equation}
where the density of the bending force is given by
\begin{multline}\label{SI:EQ:forces}
    \bm{\mathfrak{f}}_\mathrm{b}=\bigg[-\frac{\partial \mathcal{E}_\mathrm{b}}{\partial x\left(\xi\right)}+\frac{\dd}{\dd\xi}\frac{\partial \mathcal{E}_\mathrm{b}}{\partial \dot{x}\left(\xi\right)}-\frac{\dd^2}{\dd\xi^2}\frac{\partial \mathcal{E}_\mathrm{b}}{\partial \ddot{x}\left(\xi\right)}+\ldots,\\
    -\frac{\partial \mathcal{E}_\mathrm{b}}{\partial y\left(\xi\right)}+\frac{\dd}{\dd\xi}\frac{\partial \mathcal{E}_\mathrm{b}}{\partial \dot{y}\left(\xi\right)}-\frac{\dd^2}{\dd\xi^2}\frac{\partial \mathcal{E}_\mathrm{b}}{\partial \ddot{y}\left(\xi\right)}+\ldots\bigg],
\end{multline}
and the other two densities are similarly calculated. Since in the formulae for the densities of energies, at most the second derivatives of $x\left(\xi\right)$ and $y\left(\xi\right)$ appear, the terms omitted in Eq.~\eqref{SI:EQ:forces} do not produce any terms in the forces. However, a fourth derivative of $x\left(\xi\right)$ and $y\left(\xi\right)$ appears in the formula for $\bm{\mathfrak{f}}_\mathrm{b}$.

The formulae for the forces can be calculated straightforwardly, and the results are quite complicated; we refrain from presenting them here.

We assume that the motion of the membrane is overdamped with the net acting force balanced with friction proportional to the local velocity $\bm{v}\left(\xi\right)$. The motion of a membrane element $\dd \xi$ is given by 
\begin{equation}
    \bm{\mathfrak{f}}_\mathrm{b}\left(\xi\right)\dd \xi+\bm{\mathfrak{f}}_\mathrm{st}\left(\xi\right)\dd \xi+\bm{\mathfrak{f}}_\mathrm{int}\left(\xi\right)\dd \xi=\bm{v}\left(\xi\right)\gamma_\mathrm{mem} \dd \ell,
\end{equation}
where $\gamma_\mathrm{mem}$ is the friction coefficient and $\dd \ell=\left[\dot{x}^2\left(\xi\right)+\dot{y}^2\left(\xi\right)\right]^{1/2}\dd\xi$ is the length of the element $\dd\xi$. Using the approximate relation (Eq.~\eqref{SI:EQ:dl}), we get
\begin{equation}\label{SI:EQ:band_dynamics}
    \bm{\mathfrak{f}}_\mathrm{b}\left(\xi\right)+\bm{\mathfrak{f}}_\mathrm{st}\left(\xi\right)+\bm{\mathfrak{f}}_\mathrm{int}\left(\xi\right)=\bm{v}\left(\xi\right)\gamma_\mathrm{mem} \frac{\ell_0}{2\pi},
\end{equation}
which allows us to calculate the velocity of the membrane.

\subsubsection{Membrane length conservation and relaxation \\time scale}

In the simulations, when there is no interaction between the bot and the membrane, the membrane relaxes to an equilibrium circular shape due to frictional dissipation. This circular shape can be uniformly parametrized with
\begin{equation}
    x\left(\xi\right)=R \cos\xi, \qquad y\left(\xi\right)=R\sin\xi,
\end{equation}
where $R$ is the radius, the equilibrium value of which has yet to be determined. Using Eq.~\eqref{SI:EQ:forces} we have calculated
\begin{equation}\label{SI:EQ:circle_forces}
    \bm{\mathfrak{f}}_\mathrm{b}\left(\xi\right)=\frac{k}{R^2}\hat{\bm{n}}\left(\xi\right), \qquad \bm{\mathfrak{f}}_\mathrm{st}\left(\xi\right)=2\alpha \ell_0\left(\ell_0-2\pi R\right)\hat{\bm{n}}\left(\xi\right),
\end{equation}
where $\hat{\bm{n}}\left(\xi\right)$ is a normal unit vector of the membrane. The membrane is in equilibrium when $\bm{\mathfrak{f}}_\mathrm{b}+\bm{\mathfrak{f}}_\mathrm{st}=\bm{0}$, which leads to the relation
\begin{equation}
    4\pi\alpha \ell_0 R^3-2\alpha\ell_0 R^2-k=0,
\end{equation}
and the solution
\begin{equation}
    R=\frac{\ell_0}{2\pi}\left[1+2\pi^2 \frac{k}{\alpha \ell_0^4}+\mathrm{O}\left[\left(\frac{k}{\alpha \ell_0^4}\right)^2\right]\right].
\end{equation}
This shows that the length of the membrane $2\pi R$ is close to the desired length $\ell_0$ only when 
\begin{equation}\label{SI:EQ:small_k}
    k\ll \alpha \ell_0^4,    
\end{equation}
\ie, when surface tension effects dominate over elastic ones. The study of the system beyond Eq.~\eqref{SI:EQ:small_k}, where the elastic forces are strong enough to stretch the membrane significantly, is outside the scope of the current manuscript. We note that Eq.~\eqref{SI:EQ:small_k} ensures the validity of the approximation (Eq.~\eqref{SI:EQ:dl}).

When $R=\ell_0/\left(2\pi\right)+\delta R$, with $\delta R\ll R$, using Eqs.~\eqref{SI:EQ:circle_forces} and~\eqref{SI:EQ:small_k}, and from Eq.~\eqref{SI:EQ:band_dynamics}, we derive
\begin{equation}
    \bm{v}\left(\xi\right) \hat{\bm{n}}\left(\xi\right)=\frac{\dd \delta R}{\dd t}=-\frac{8\pi^2\alpha}{\gamma_\mathrm{mem}}\delta R,
\end{equation}
where $t$ denotes time.
This allows us to identify the timescale
\begin{equation}
    t_\mathrm{mem}=\frac{\gamma_\mathrm{mem}}{8\pi^2 \alpha}
\end{equation}
which characterizes the process of reaching by the membrane its equilibrium shape. 

\subsection{Modeling the bot}

We assume that the bot has an elliptical shape with the same length scales as the experiments. The geometric center of the bot is located at $\bm{r}_\mathrm{bot}=\left[x_\mathrm{bot},y_\mathrm{bot}\right]$ on a two-dimensional plane, and the tilt angle between the longer axis and the $x$-direction is given by $\vartheta_\mathrm{bot}$.

\subsubsection{Time evolution of the bot motion}

The total force $\bm{\mathfrak{F}}_\mathrm{bot}$ and torque $\mathfrak{T}_\mathrm{bot}$ acting on the bot can naturally be decomposed into two parts
\begin{equation}
    \bm{\mathfrak{F}}_\mathrm{bot}=\bm{\mathfrak{F}}_\mathrm{int-bot}+\bm{\mathfrak{F}}_\mathrm{dr}, \qquad \mathfrak{T}_\mathrm{bot}=\mathfrak{T}_\mathrm{int-bot}+\mathfrak{T}_\mathrm{dr},
\end{equation}
originating from the interaction with the membrane (index ``$\mathrm{int-bot}$'') and driving protocol (index ``$\mathrm{dr}$''), such as Brownian stochastic noise or active processes.

The driving protocol is discussed in Sec.~S2B2. The force and torque of the interaction with the membrane can be derived straightforwardly from the energy of this interaction (Eq.~\eqref{SI:EQ:Eint}) as
\begin{subequations}\label{SI:EQ:int_bot}
\begin{align}
    \bm{\mathfrak{F}}_\mathrm{int-bot}&=-\int_0^{2\pi}\dd\xi\,\bm{\mathfrak{f}}_\mathrm{int}\left(\xi\right),\\
    \mathfrak{T}_\mathrm{int-bot}&=-\int_0^{2\pi}\dd\xi\,\bm{\gamma}\left(\xi\right)\times\bm{\mathfrak{f}}_\mathrm{int}\left(\xi\right).\label{SI:EQ:T_int_bot}
\end{align}
\end{subequations}
We note that while the forces are considered here to be two-dimensional vectors, the torque is a scalar quantity, with its positive value meaning rotation of the bot anticlockwise. The vector product present in Eq.~\eqref{SI:EQ:T_int_bot} should be interpreted as an operation on three-dimensional vectors created by adding a zero $z$-th component to two two-dimensional vectors, and the $z$-th component of the resulting vector is taken as the final result. 

Similarly to the membrane, we assume overdamped dynamics of the bot with the forces and torques balanced by the friction effects proportional to the velocities
\begin{equation}\label{SI:EQ:bot_dynamics}
    \bm{\mathfrak{F}}_\mathrm{bot}=\gamma_\mathrm{bot}^\mathrm{T}\frac{\dd \bm{r}_\mathrm{bot}}{\dd t}, \qquad \mathfrak{T}_\mathrm{bot}=\gamma_\mathrm{bot}^\mathrm{R}\frac{\dd \vartheta_\mathrm{bot}}{\dd t},
\end{equation}
where $\gamma_\mathrm{bot}^\mathrm{T}$ and $\gamma_\mathrm{bot}^\mathrm{R}$ denote the translational and rotational damping coefficients of the bot, respectively.

\begin{figure}[t!]
    \centering
    \includegraphics[width=\linewidth]{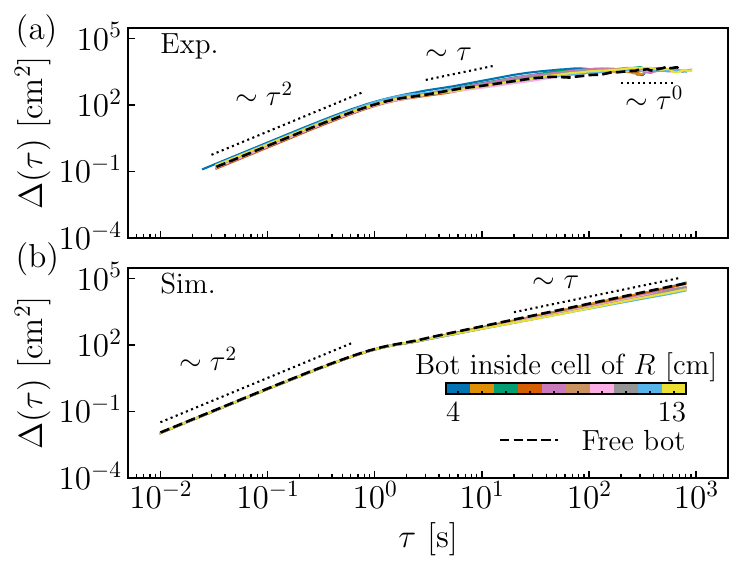}
    \caption{Mean-Squared-Displacement $\Delta$ against time lag $\tau$ for the free bot (bot in the absence of the membrane) and the bot inside the membrane for (a) experiments, and (b) simulations.}
    \label{fig:msd_bot}
\end{figure}

\subsubsection{Transport properties of the bot}\label{subsubsec:bot_driving}

In the simulations, the bot translates in the direction of its heading with a constant force $\bm{\mathfrak{F}}_\mathrm{dr}$ for a duration $\tau_\mathrm{r}$. The translational motion is then followed by a tumble phase, characterized by the rotation of the bot about its geometric center with a constant torque $\bm{\mathfrak{T}}_\mathrm{dr}$ for a finite duration $\tau_\mathrm{t}$. The run-times $\tau_\mathrm{r}$ and the tumble-times $\tau_\mathrm{t}$ are sampled from a uniform distribution $\mathcal{U}[\tau_\mathrm{l}, \tau_\mathrm{h}]$. In each tumble phase, the bot is equally likely to rotate either counterclockwise or clockwise. 

The resulting Mean-Squared Displacement (MSD) of such a bot in open space without membrane constraints consists of a ballistic regime on short time scales, which crosses over to a diffusive regime at long time scales ('Free bot' in Fig.~\ref{fig:msd_bot}(b)). On the contrary, placing a bot in the experimental domain of $1$ m$\times 1$ m results in a short diffusive regime at intermediate values, with finite size effects dominating long time dynamics (Fig.~\ref{fig:msd_bot}(a)). The persistence time of the bot in the experimental arena has been calculated by fitting $\Delta_c \simeq v_\mathrm{bot}\,\tau^2$ and $\Delta_c \simeq D_\mathrm{bot}\,\tau$ to the ballistic and diffusive regimes, respectively, and finding the crossover of the two curves with an uncertainty arising from that of the fitting parameters. As the bot behaves like an effective Brownian particle at long time scales, the diffusion coefficient is calculated by fitting the MSD of the free bot in the experiments to the expression for the MSD of a Brownian particle in a 2D confined square domain [37], considering the first $51$ terms of the power series (Eq.~\eqref{SI:EQ:MSD_confined}). Increasing the number of power series terms did not affect the $D_\mathrm{bot}$ to the order of $\mathcal{O}(10^{-3})$. 
\begin{multline} \label{SI:EQ:MSD_confined}
    \Delta_c(\tau) = \frac{L^2}{3} - \frac{32L^2}{\pi^4}\sum_{p=0}^{50}\frac{1}{\left(2p+1\right)^4} e^{-\frac{\pi^2\,D_\mathrm{bot}\,\tau}{L^2}\left(2p+1\right)^2} 
\end{multline}

Even when the bot is confined to move inside the membrane, the MSD of the bot remains almost unaffected (Fig.~\ref{fig:msd_bot}).

\subsection{Simulation algorithm}
To perform numerical calculations of the time evolution of the system, we describe the membrane using $N$ points homogeneously distributed in the parameter domain
\begin{equation}
    \bm{\gamma}_n^{(t)}=\left[x_n^{(t)},y_n^{(t)}\right]=\left[x\left(\frac{2\pi n}{N}\right), y\left(\frac{2\pi n}{N}\right)\right],
\end{equation}
with $n=0,1,2,\ldots,N-1$, and $t$ denoting the simulation time. The bot is described with its position $\bm{r}_\mathrm{bot}^{(t)}=\left[x_\mathrm{bot}^{(t)}, y_\mathrm{bot}^{(t)}\right]$ and orientation $\vartheta_\mathrm{bot}^{(t)}$. These parameters are evolved in time following the dynamics described in the previous section. The numerical value of $N$, as well as all other parameters, is listed in Table S1.

The algorithm consists of the following steps:
\begin{enumerate}
    \item We first derive the Fourier coefficients $g_k$
    \begin{equation}
        x_n^{(t)}+\iu y_n^{(t)}=\sum_k g_k \exp\left(\iu k\frac{2\pi  n }{N}\right),
    \end{equation}
where $k=-N/2+1,-N/2+2,\ldots, N/2$, and $\iu$ denotes the imaginary unit. This calculation is done using the fast Fourier transform implemented in the GSL library [38].
\item We approximate the shape of the membrane by
\begin{equation}\label{SI:EQ:shape_approx}
    \bar{x}\left(\xi\right)+\iu \bar{y}\left(\xi\right)=\sum_{k=-\Lambda_\mathrm{c}}^{\Lambda_\mathrm{c}} g_k \exp\left(\iu k \xi\right),
\end{equation}
where $\Lambda_\mathrm{c}$ is the cutoff parameter. The above formula is also used to estimate the derivatives of $x\left(\xi\right)$ and $y\left(\xi\right)$.
\item Using Eq.~\eqref{SI:EQ:shape_approx} we calculate estimates of the forces $\bar{\bm{\mathfrak{f}}}_\mathrm{b}\left(2\pi n/N\right)$, $\bar{\bm{\mathfrak{f}}}_\mathrm{st}\left(2\pi n/N\right)$, and $\bar{\bm{\mathfrak{f}}}_\mathrm{int}\left(2\pi n/N\right)$ for $n=0,1,2,\ldots, N-1$. The formulae for these forces were derived using Eq.~\eqref{SI:EQ:forces}.
\item We update the shape of the membrane following Eq.~\eqref{SI:EQ:band_dynamics} as
\begin{multline}\label{SI:EQ:band_numerical_evolution}
    \bm{\gamma}_n^{(t+\dd t)} =\bar{\bm{\gamma}}\left(\frac{2\pi n}{N}\right)+\frac{2\pi \dd t}{\gamma_\mathrm{mem} \ell_0}\left[\bar{\bm{\mathfrak{f}}}_\mathrm{b}\left(\frac{2\pi n}{N}\right)\right. \\
     \left.+\bar{\bm{\mathfrak{f}}}_\mathrm{st}\left(\frac{2\pi n}{N}\right)+\bar{\bm{\mathfrak{f}}}_\mathrm{int}\left(\frac{2\pi n}{N}\right)\right],
\end{multline}
where $\dd t$ is the time step, and $\bar{\bm{\gamma}}\left(\xi\right)=\left[\bar{x}\left(\xi\right),\bar{y}\left(\xi\right)\right]$. We note that Eq.~\eqref{SI:EQ:shape_approx} was used as an estimate of the shape of the membrane at time $t$ rather than $\bm{\gamma}_n^{(t)}$. This effectively removes all higher frequencies from the shape of the membrane.
\item We estimate the force and torque of interaction between the membrane and the bot using Eq.~\eqref{SI:EQ:int_bot} as
\begin{subequations}
\begin{align}
    \bar{\bm{\mathfrak{F}}}_\mathrm{int-bot}&=-\frac{2\pi}{N}\sum_{n=0}^{N-1}\bar{\bm{\mathfrak{f}}}_\mathrm{int}\left(\frac{2\pi n}{N}\right),\\
    \bar{\mathfrak{T}}_\mathrm{int-bot}&=-\frac{2\pi}{N}\sum_{n=0}^{N-1}\bar{\bm{\gamma}}\left(\frac{2\pi n}{N}\right)\times\bar{\bm{\mathfrak{f}}}_\mathrm{int}\left(\frac{2\pi n}{N}\right).   
\end{align}
\end{subequations}
\item We calculate the force and torque acting on the bot stemming from the internal effects. During the run phase of the bot's motion, the bot is subject to a constant magnitude of force $F_0$ in the direction of the current bot heading, while during the tumble phase, a constant torque $T_0$ acts on the bot with an equal probability of going clockwise or counterclockwise. 

\begin{table*}[t!]
    \centering
    \begin{tabular}{ccc}
    Parameter & Description & Value \\
    \hline
      $A$ & Amplitude of bot-membrane interaction & $0.000285$ [N]\\
      $\alpha$ & Surface tension coefficient of membrane & $0.02$ [N$\cdot$m$^{-2}$]\\
      $\mathrm{d} t$ & Time step for the simulation & $10^{-6}$ [s] \\
      $F_{0}$ & Translation force on bot & $0.04275$ [N]\\
      $\gamma_\mathrm{bot}^\mathrm{R}$ & Rotational friction coefficient of bot  & $0.285$ [N$\cdot$m$\cdot$s]\\
      $\gamma_\mathrm{bot}^\mathrm{T}$ & Translational friction coefficient of bot & $0.285$ [N$\cdot$s$\cdot$m$^{-1}$]\\
      $\gamma_\mathrm{mem}$ & Friction per unit length of membrane & $0.000342$ [N$\cdot$s$\cdot$m$^{-2}$] \\
      $k$ & Bending stiffness of membrane & $0.57$ [N$\cdot$m$^{2}$]\\
      $\lambda$ & Length scale of bot-membrane interaction & $0.0001$ [m]\\
      $N$ & Number of points on the membrane & $1024$ \\
      $T_{0}$ & Torque on bot & $1.511$ [N$\cdot$m]\\
      $\tau_\mathrm{h}$ & Upper time limit of one bot translation/rotation & $1.2$ [s] \\
      $\tau_\mathrm{l}$ & Lower time limit of one bot translation/rotation & $0.4$ [s]\\
    \hline
    \end{tabular}
    \caption{Parameters for the simulation.}
    \label{tab:params}
\end{table*}

\begin{equation}\label{SI:EQ:bot_forces}
    \bm{\mathfrak{F}}_\mathrm{dr}=F_{0} \left(\begin{array}{cc}
       \cos \vartheta\\
        \sin \vartheta \\
    \end{array}\right), \qquad \mathfrak{T}_\mathrm{dr}= \pm T_{0},
\end{equation}

\item Finally, we update the bot configuration following Eq.~\eqref{SI:EQ:bot_dynamics}:
\begin{subequations}
\begin{align}
    \bm{r}_\mathrm{bot}^{(t+\dd t)}=\bm{r}_\mathrm{bot}^{(t)}+\frac{\dd t}{\gamma_\mathrm{bot}^\mathrm{T}}\left(\bar{\bm{\mathfrak{F}}}_\mathrm{int-bot}+\bm{\mathfrak{F}}_\mathrm{dr}\right),\\
    \vartheta_\mathrm{bot}^{(t+\dd t)}=\vartheta_\mathrm{bot}^{(t)}+\frac{\dd t}{\gamma_\mathrm{bot}^\mathrm{R}}\left(\bar{\mathfrak{T}}_\mathrm{int-bot}+\mathfrak{T}_\mathrm{dr}\right).
\end{align}
\end{subequations}
\end{enumerate}
The steps presented above advance the system time by $\dd t$, and need to be repeated consecutively in the numerical simulation.

We have decided to use the Fourier transform for interpolating the shape of the membrane from the stored points $\bm{\gamma}_n^{(t)}$ because this approach generates the shape as a smooth periodic function with well-defined derivatives, even of higher orders. As we have checked, the time evolution of higher frequencies in the spectrum of the shape of the membrane is generally less numerically stable and requires a smaller value of $\dd t$ to produce reliable results. That is why we have introduced the cutoff $\Lambda_\mathrm{c}$ in Eq.~\eqref{SI:EQ:shape_approx}. Finally, in Eq.~\eqref{SI:EQ:band_numerical_evolution} we have decided to use the shape of the membrane $\bar{\bm{\gamma}}\left(\xi\right)$ after the cutoff as a base for the time evolution rather than the original shape $\bm{\gamma}_n^{(t)}$. This was done to remove all the higher frequencies from the spectrum of the shape of the membrane---without this, these frequencies are completely free parameters (as they do not contribute to the forces) and, in time, they collect numerical noise, making the membrane highly corrugated.

\subsection{Parameters}
The bot-related parameters are chosen to mimic the bot kinematics from the experiments with the design specifications of a Graspion [32]. The translational friction coefficient acting on the bot $\gamma_\mathrm{bot}^\mathrm{T}$ has also been measured from experimental observations. The bot has been measured to exert a force of $42.75$ mN while moving at a constant speed of $15$ cm$\cdot$s$^{-1}$, and assuming overdamped dynamics, we obtain $\gamma_\mathrm{bot}^\mathrm{T}=F/v\simeq 0.285$. On the other hand, the rotational friction coefficient $\gamma_\mathrm{bot}^\mathrm{R}$ has been set such that the sliding of the bot along the membrane occurs rarely.

\section{Data processing}

The membrane is discretized into $N=1024$ points, which is observed to be large enough to capture all experimental shape features for the largest cell size ($R=13$ cm). Among other membrane-related parameters, the friction per unit length $\gamma_\mathrm{mem}$ has been chosen from preliminary simulations, where the bot continuously pushes the membrane along a fixed direction. The temporally averaged absolute value of the velocity of the cell's geometric center is matched against that obtained from similar experiments. The surface tension coefficient $\alpha$ is chosen sufficiently high, so that the change in the length of the membrane during the simulations never exceeds $\pm0.1\%$. The bending stiffness $k$ is selected by comparing the deformability of the membrane in the experiments and the simulations. This is quantified in terms of a dimensionless shape index $\mathrm{IPR}=P_\mathrm{mem}/\sqrt{A_\mathrm{mem}}$, where $P_\mathrm{mem}$ and $A_\mathrm{mem}$ are the perimeter and the area of the deformed membrane, respectively. We vary $k$ over a range of values until we obtain a close match with the temporally averaged IPR from the experiments for all cell sizes.

The parameters corresponding to the bot-membrane interaction $A$ and $\lambda$ are selected such that there is no overlap between the two objects during the simulations.

\subsection{Experimental data analysis} \label{subsec:exp_data_analysis}
The positions of both the bot and the geometric center of the membrane are used to compute the velocity as a function of time and the mean-squared displacement (MSD)
\begin{equation}\label{EQ:MSD}
    \Delta(\tau) = \left\langle \left|\bm{r}_\mathrm{mem}^{(t+\tau)} - \bm{r}_\mathrm{mem}^{(t)} \right|^2 \right\rangle
\end{equation}
\noindent where $\tau$ is the time lag, $\bm{r}_\mathrm{mem}^{(t)}$ is the position vector of the membrane's geometric center at time $t$, and $\langle \cdot \rangle$ denotes averaging over all data collected across realizations for a given time lag. An analogous calculation was performed for the bot MSD, using $\bm{r}_\mathrm{bot}^{(t)}$ as the bot position vector at time $t$.

The MSDs are then fitted with power-law expressions with exponents $2$ and $1$, corresponding to ballistic and diffusive regimes, respectively, in order to characterize the transport properties of the membrane. 

The velocities of both the membrane and the bot are computed from the first temporal derivative of their respective positions. The stop and the go phases of the membrane motion are then identified using velocity thresholds, set at $2\%$ of the average bot velocity (i.e., $0.3$ cm$\cdot$s$^{-1}$), determined based on the velocity time series and observed noise in the stop phases. Consecutive time steps with velocities below the threshold define stop phases, whereas those above the threshold define go phases. The duration of each phase is computed as the time interval between the first and last time step of that sequence. Varying the threshold by $\pm 0.1$ cm$\cdot$s$^{-1}$ does not alter the distributions or the mean value of the stop and go times significantly.

The probability density functions (PDFs) of the stop and go durations are then constructed, for each $R$, from the full set of measured durations across the experiments. The durations obtained for different $R$ are rescaled with their respective mean values as $t'=t/\langle t\rangle$, and the PDFs are plotted in the form of histograms of constant bin size = $0.25$. The collapse of the PDFs onto a single master curve indicates that the stop and the go durations are governed by the same underlying exponential process $p(t')=e^{-t'}$, with a difference only in the characteristic timescales. These characteristic stop and go times are obtained from the unscaled duration data, and denoted $\langle\tau_\mathrm{s}\rangle$ and $\langle\tau_\mathrm{g}\rangle$, respectively.

\begin{figure}[t!]
    \centering
    \includegraphics[width=\linewidth]{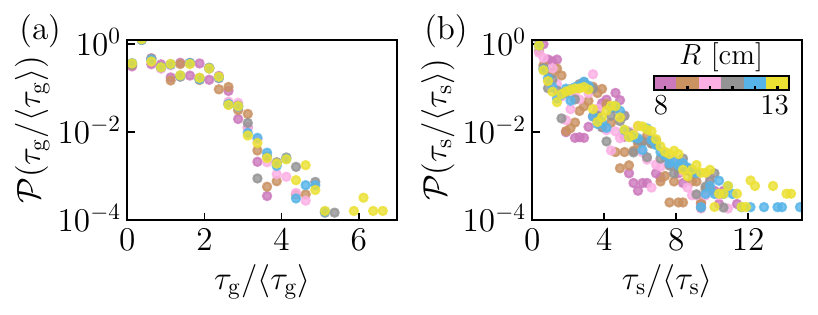}
    \caption{PDFs $\mathcal{P}$($\cdot$) of (a) go-times $\tau_\mathrm{g}$ and (b) stop times $\tau_\mathrm{s}$ for different cell sizes, normalized by the respective mean. $\circ$ represents the heights of the bins of the histograms.}
    \label{fig:simul_pdf_stop_go}
\end{figure}

\subsection{Simulation data analysis}
Twenty-five realizations of the simulations are performed for each cell size $R \in [4, 13]$, using the parameter set mentioned in Table~\ref{tab:params} on an in-house routine in C++. The position of the geometric center of the bot $\bm{r}_\mathrm{bot}$, along with the orientation $\vartheta_\mathrm{bot}$, as well as the positions of all the points on the membrane $\gamma_n \quad \forall\quad n = 0,1,2\ldots N-1$, are stored every $10^4\dd t$ for the first $100$ s, and every $10^5 \dd t$ thereafter. Next, the geometric center (GC) of the membrane $\bm{r}_\mathrm{mem}$ is calculated for all time instances by an algebraic average of the membrane points.

Next, the velocity of the membrane GC, $v(t)$, is computed by performing a first-order forward difference of the membrane GC positions. The time series of the absolute value of the velocity is then filtered with a minute threshold of $|v|<0.03$ cm$\cdot$s$^{-1}$ to demarcate the stop instances. All the time instances where $|v|>0.03$ cm$\cdot$s$^{-1}$ are marked as part of a go-phase. This threshold is important so that minute velocities emerging from numerical errors don't get classified as a go-phase of the cell. The above process is repeated for all realizations across cell sizes, and the durations of the go and the stop phases are collated. These data have been presented in the form of PDFs after normalization with their respective means in Fig.~\ref{fig:simul_pdf_stop_go}, similar to Sec.~S3A. The circles represent the height of the bins of the histograms of constant bin size = $0.25$. 

\begin{figure}[t!]
    \centering
    \includegraphics[width=0.99\columnwidth]{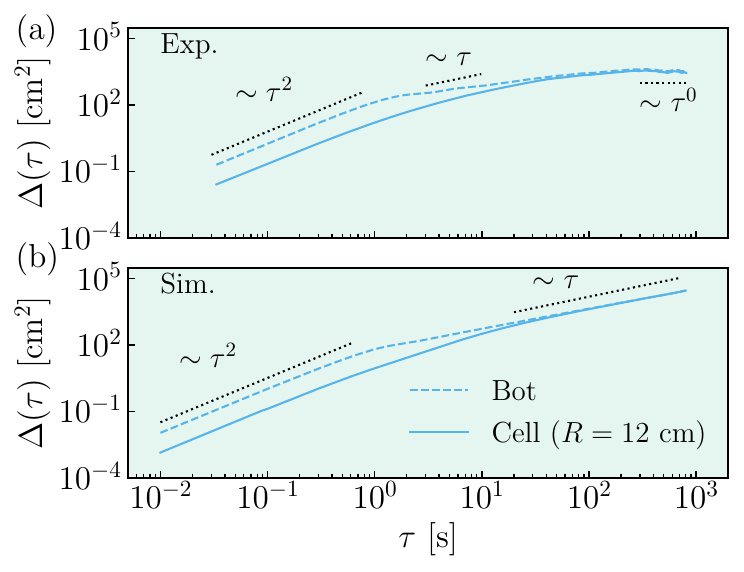}
    \caption{Mean-Squared Displacement for a cell of radius $R=12$ cm, and the bot placed inside the cell, in the case of (a) experiments, and (b) simulations. The green shading represents the decoupled dynamical state of the cell.}
    \label{fig:msd_R12}
\end{figure}

The Mean-Squared Displacement of the GCs of the bot and the membrane were also calculated from their respective position data using Eq.~\eqref{EQ:MSD}. The positions from the first $15$ s were discarded during this computation, to avoid inclusion of initial transience. While the MSDs of the bot and the cell follow distinct ballistic and crossover regimes, they merge on long time scales, indicating that the two move as a single diffusive entity (Fig.~\ref{fig:msd_R12}). This is expected once the transients related to deformation and persistence have decorrelated. 

To compute the ballistic velocity and the diffusion coefficient from the MSD, a procedure similar to that mentioned in Sec.~S3A was used.

\section{Theoretical MSD of the cell}

The mean-squared displacement can be expressed in terms of the velocity autocorrelation function (VACF) as
\begin{equation}\label{SI:EQ:theo_msd}
    \Delta(\tau) = \langle |\bm{r}_\mathrm{mem}^{(\tau)} - \bm{r}_\mathrm{mem}^{(0)}|^2 \rangle = \int_{0}^{\tau} \dd s \int_{0}^{\tau} \langle\mathbf{v}(s)\cdot \mathbf{v}(s') \rangle \dd s',
\end{equation}
where $\bm{r}_\mathrm{mem}^{(\tau)}$ is the position vector of the membrane's geometric center at time $\tau$, and $s$ as well as $s'$ are two arbitrary time instances, and $\mathbf{v}(s)$ represents instantaneous velocity at time $s$. For a stationary stochastic process, the VACF depends only on the time lag $z=|s-s'|$. So, Eq.~\eqref{SI:EQ:theo_msd} can be rewritten as
\begin{equation}\label{SI:EQ:theo_msd_stationary}
    \Delta(\tau) = 2\int_{0}^{\tau} (\tau-z)\,\langle\mathbf{v}(0)\cdot \mathbf{v}(z) \rangle \dd z.
\end{equation}

In the current work, the cell exhibits stop-and-go motion with alternating stop and go phases. In the intermediate and decoupled dynamical states, these stop and go durations are exponentially distributed with mean values $\langle\tau_\mathrm{s}\rangle$ and $\langle\tau_\mathrm{g}\rangle$, respectively. Therefore, we treat the motion as a run-and-tumble process, with runs corresponding to go phases, and tumbles corresponding to stop phases. During a run, the cell moves with a constant velocity $v\, \hat{\mathbf{n}}$, where $v$ is the run speed and $\hat{\mathbf{n}}$ is a unit vector specifying the direction of motion, while during a stop phase, there is no translation of the cell ($v=0$). The run speed is sampled from a stochastic process with finite moments. We denote $\bar{v}$ and $\sigma_v^2$ as the mean and the variance of the full speed distribution measured over the entire time series, including the zero-velocity stop events. 

To obtain the VACF, we impose the following assumptions. The angular diffusion during a run is neglected, so the orientation remains fixed throughout an individual run event. Secondly, the thermal diffusion is assumed to be negligible in both the go and the stop phases. Furthermore, each tumbling (stop event) fully randomizes the direction of the subsequent run, so the VACF vanishes if the times $0$ and $z$ lie in different go events. The VACF also vanishes when either of $0$ or $z$ lies in the stop phase, as $v=0$. The scalar product $\mathbf{v}(0)\cdot\mathbf{v}(z)$ is non-zero only when $0$ and $z$ lie in the same run segment. In such a case, the VACF reduces to 
\begin{equation}\label{SI:EQ:vacf_part}
    \langle\mathbf{v}(0)\cdot \mathbf{v}(z) \rangle = \langle v^2\rangle \,\langle\hat{\mathbf{n}}(0)\cdot\hat{\mathbf{n}}(z)\rangle,
\end{equation}
where $\langle v^2 \rangle = \bar{v}^2 + \sigma_v^2$ denotes the second moment of the full velocity distribution, and the scalar product $\langle\hat{\mathbf{n}}(0)\cdot\hat{\mathbf{n}}(z)\rangle$ represents the orientational auto-correlation. As the go (run) durations are exponentially distributed, the probability that a run starting at time $0$ survives without termination up till time $z$ is $e^{-z/\langle\tau_\mathrm{g}\rangle}$. Since $\langle v^2 \rangle$ is defined here, with respect to the full velocity distribution, the stationary probability of a run phase is already absorbed into this quantity, and no additional prefactor $\langle \tau_\mathrm{g}\rangle / \left(\langle \tau_\mathrm{g}\rangle + \langle \tau_\mathrm{s}\rangle \right)$ is required. Consequently, the VACF takes the form
\begin{equation}\label{SI:EQ:vacf}
    \langle\mathbf{v}(0)\cdot \mathbf{v}(z) \rangle = \left(\bar{v}^2 + \sigma_v^2 \right) \, e^{-z/\langle \tau_\mathrm{g}\rangle}.
\end{equation}
On substituting Eq.~\eqref{SI:EQ:vacf} into Eq.~\eqref{SI:EQ:theo_msd_stationary}, and integrating it, we obtain the MSD as
\begin{equation}\label{SI:EQ:msd}
    \Delta(\tau) = 2(\bar{v}^2 + \sigma_v^2) \left(\tau \langle \tau_\mathrm{g}\rangle - \langle \tau_\mathrm{g}\rangle^2 \left(1 - e^{-\tau/\langle \tau_\mathrm{g}\rangle} \right)\right).
\end{equation}
For short time scales ($\tau \ll \langle \tau_\mathrm{g}\rangle$), the above expression reduces to
\begin{equation}
    \Delta(\tau) \simeq \left(\bar{v}^2 + \sigma_v^2 \right)\, \tau^2,
\end{equation}
where $\sqrt{\bar{v}^2 + \sigma_v^2}$ is the ballistic velocity.

On long time scales ($\tau\gg\langle \tau_\mathrm{g}\rangle$), the dynamics become asymptotically diffusive as
\begin{equation}
    \Delta(\tau) \simeq 2\left(\bar{v}^2 + \sigma_v^2\right)\,\langle \tau_\mathrm{g}\rangle \,\tau,
\end{equation}
where $2\left(\bar{v}^2 + \sigma_v^2\right)\,\langle \tau_\mathrm{g}\rangle$ is the effective diffusion coefficient of the cell.

\end{document}